\documentclass[aps,prl,groupedaddress,twocolumn,showpacs,
nofootinbib]{revtex4}  
\usepackage{graphicx}  
\usepackage{dcolumn}   
\usepackage{bm}        
\usepackage{amssymb}   
\usepackage{amsmath}
\usepackage{mdframed}
\usepackage{theorem}
\usepackage{tikz}
\usetikzlibrary{trees}
\usetikzlibrary{decorations.pathmorphing}
\usepackage{bbm}
\usepackage{here}

\newcommand{\ket}[1]{\ensuremath{|#1\rangle}}

\newcommand{\bra}[1]{\ensuremath{\langle#1|}}

\newcommand{\EE}{\ensuremath{\mathcal{E}}}
\DeclareMathOperator{\BB}{\mathcal{B}}
\DeclareMathOperator{\vrho}{\varrho}
\DeclareMathOperator{\II}{\mathcal{I}}
\newcommand{\bea}{\begin{eqnarray}}
\newcommand{\eea}{\end{eqnarray}}
\newcommand{\tr}{\ensuremath{{\rm tr}}}

\tikzset{
  level/.style   = {line width=1.5pt},
  connect/.style = { dashed },
  transition/.style   = {line width=0.5pt},
  decay/.style={decorate, decoration={snake}, draw=red}
}
\definecolor{cobalt}{RGB}{61, 82, 200}
\definecolor{lila}{RGB}{127, 0, 127}

\begin{document}
\title{Genuine temporal correlations can certify the quantum dimension}

\author{Cornelia Spee}
\author{Hendrik Siebeneich}
\author{Timm Florian Gloger}
\author{Peter Kaufmann}
\author{ Michael Johanning}
\author{ Matthias Kleinmann}
\author{Christof Wunderlich}
\author{Otfried G\"uhne}

\affiliation{Naturwissenschaftlich-Technische Fakult\"at, 
Universit\"at Siegen, Walter-Flex-Stra{\ss}e 3, 57068 Siegen, Germany}

\date{\today}


\begin{abstract}
Temporal correlations in quantum mechanics are the origin
of several non-classical phenomena, but they depend on the dimension of 
the underlying quantum system. This allows one to use such correlations for the
certification of a minimal Hilbert space dimension. Here we provide a theoretical proposal and an experimental implementation of a device-independent dimension test, using temporal correlations observed on a single 
trapped $^{171}$Yb$^+$ ion. Our test goes beyond the prepare-and-measure scheme of 
previous approaches, demonstrating the advantage of genuine temporal correlations.
\end{abstract}
\maketitle


{\it Introduction.---}
Correlations arising from sequential measurements on a single system can be used to construct  Kochen-Specker \cite{KS} and Leggett-Garg inequalities \cite{LeggettGarg}, which test whether the system's dynamics and measurements admit a classical description. In particular, Leggett-Garg inequalities hold if a theory fulfills the assumptions of  macrorealism and non-invasive measurability. Quantum mechanics does not satisfy these conditions and violations of Leggett-Garg inequalities have been observed in experiments \cite{LeggettGargExp1,LeggettGargExp2}. Moreover, noncontextuality inequalities have been experimentally violated (see e.g. \cite{ContextExp1}).  This triggered research on the question which temporal correlations can be realized within quantum mechanics \cite{tc0,tc1,tc2,tc3,tc4}. For the spatial scenario, i.e., the Bell scenario, it is well known that there exist no-signaling correlations that cannot be obtained within quantum mechanics \cite{PRboxes}. In contrast to that, for the temporal scenario it is possible to obtain all correlations that do not exhibit signaling with respect to the past within quantum theory, if one does not restrict the dimension of the quantum system and the type of measurements \cite{Fritz,temporal}. If the dimension of the system is restricted, however, some correlations cannot be realized \cite{temporal}. This allows one to exploit temporal correlations for testing the dimension of a quantum system. Sequential measurements can also be used to witness quantum coherence \cite{witquantcoh}.

Certifying the minimal quantum dimension is an important task for the following reasons. First, it has  been realized that for some applications in quantum information theory such as quantum key distribution  high-dimensional systems are advantageous compared to low-dimensional ones \cite{QKD1, QKD2}. Second, high-dimensional systems became within reach of current technology, e.g., in photonic systems \cite{expHd0,expHd1,expHd2,expHd3,expHd4}. This requires the certification of the dimension of the system that can be accessed and manipulated in an experiment. 

Dimension witnesses are inequalities that hold true for a maximal dimension and therefore their violation provides a lower bound on the dimension. They have been put forward for different scenarios. Some of them rely on assumptions about the type of measurements \cite{Fritz,dimtemp2,witcontext,vioquantwiteq} such as their projective nature or the fact that the time evolution of the system should be (at least on a coarse-grained time scale) Markovian \cite{witdavid, dwit} or only reversible transformations are applied \cite{witgame}.

Device-independent dimension witnesses have been obtained for bipartite systems \cite{witBell} using Bell inequalities and for single systems \cite{prepmeas,prepmeas2,vio1,vio2,prepmeas3,prepmeasnetw} for the prepare-and-measure (P\&M) scenario. In this scenario states chosen from a set of states $\{\rho_\xi\}$ are prepared and then a measurement chosen out of a set of  measurements $\{\mathcal{M}_\mu\}$ is performed resulting in correlations $p(a|\xi \mu)$ for obtaining outcome $a$ given the respective state preparation $\xi$ and measurement process $\mu$. Dimension witnesses in the P\&M scenario have been experimentally implemented  \cite{vio2,exp1,exp2}. Moreover, dimensions bounds have been obtained by making use of random access codes \cite{prepmeas0}.

\begin{figure}[t!]
\begin{center}
\includegraphics[width=0.95\columnwidth]{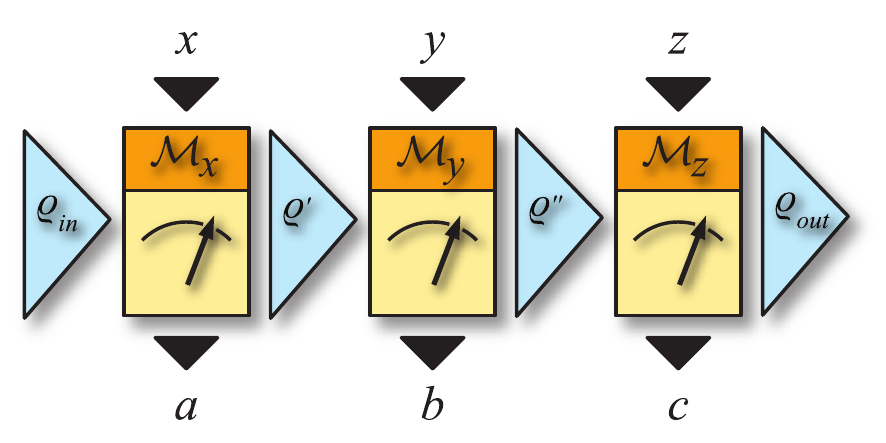} 
\end{center}
\caption{Sequences of general measurements on a quantum state $\vrho_{\rm in}$ result in correlations $p(abc\ldots|xyz\ldots)$, where $x, y, z \ldots$ denotes the measurement settings and $a, b, c  \ldots$ the respective outcomes. The measurements that are performed in the respective time steps are denoted by $\mathcal{M}_x$, $\mathcal{M}_y$ and $\mathcal{M}_z$ and the corresponding post-measurement states are given by $\varrho', \varrho''$ and  $\varrho_{\rm out}$. The same measurement device can be used at different time steps.}
\label{Fig1}
\end{figure}

Recently, dimension witnesses based on sequences of general measurements have been introduced \cite{temporal}. In this scenario sequences of  measurements are performed on a quantum state $\vrho_{\rm in}$ (see Fig. \ref{Fig1}). In particular, the most general measurements  within quantum mechanics are considered  and an arbitrary (outcome-dependent and time-dependent) transformation of the system between the measurements is allowed. Hence, the post-measurement states can depend on the initial state, previous measurement outcomes, previous choices of measurement settings and the time evolution. Note that in contrast to the P\&M scenario the post-measurement states are here essential for the correlations $p(abc\ldots|xyz\ldots)$.
The witnesses obtained in this scenario are device-independent, only relying on the assumption that at each time step the measurements are chosen out of a fixed set of measurements that does not vary with time. That is, in contrast to previous dimension witnesses using temporal correlations \cite{Fritz,dimtemp2,witcontext,vioquantwiteq,witdavid,dwit,witgame} no assumptions on the measurements or the transformations in-between the measurements have to be made.  A scenario similar to ours in which however no  relation among the measurements performed at different times is assumed has been discussed in \cite{prepmeasnetw}. But dimension bounds have been derived only based on the correlations obtained for measurement sequences of length two  \cite{temporal,prepmeasnetw}. Measurement sequences of length two may still be interpreted in the P\&M scenario. Here the preparation process would be implemented by the first measurement, whereas the second measurement would act as the actual measurement process providing the correlations. Note, however, that dimension witnesses based on temporal
inequalities may also include correlations with the measurement
outcome of the first measurement. Since measurements
can be repeated in our scenario, such correlations lack a clear
interpretation in the P\&M scenario.

In this paper we propose a dimension witness using  true multipoint temporal correlations which  does not allow for an interpretation in the P\&M scheme and shows a potentially larger ratio of violation than the dimension witnesses obtained in \cite{temporal}. We report the implementation of this witness  on a single trapped $^{171}$Yb$^+$ ion and certify the manipulation of a three-dimensional system in the experiment. For reference, the dimension witnesses of Ref. \cite{temporal} have been experimentally implemented. This shows that dimension tests based on temporal correlations are of practical relevance and can be used to certify lower bounds on the dimension with current state technology.

{\it Temporal inequalities.---}
Measurement sequences on a quantum system fulfill the arrow of time (AoT) constraints \cite{AoT}, i.e.,  the choices of the measurement settings at all subsequent time steps cannot influence the probabilities obtained for measurements that have been already implemented. The AoT constraints define for every length, number of measurements settings and number of outcomes the temporal correlation polytope. The extreme points of this polytope correspond to the deterministic assignments, i.e., the correlations are either zero or one, that obey the AoT constraints  \cite{temporal,polytope} and all correlations of this polytope can be obtained within quantum mechanics (even if measurement can be repeated) if the dimension of the quantum system is unrestricted \cite{Fritz,temporal}. However, already for the most simple scenario of sequences of length two and two measurement settings with two outcomes each there are extreme points of the polytope that cannot be reached with general  measurements on a qubit.

This dimension dependence can be used to construct temporal inequalities that act as a dimension witness. Recall that by $p(ab|xy)$ we denote the correlations obtained by performing in the first  step measurement $x\in\{0,1\}$ and obtaining outcome $a\in\{+,-\}$ and in the second time step measurement $y\in\{0,1\}$ with outcome $b\in\{+,-\}$ (see also Fig. \ref{Fig1}). Let us then consider the four sums of correlations
\begin{align}
\BB_{1}&= p(\tiny{++}|00) +p(\tiny{++}|11) + p(\tiny{+-}|01) + p(\tiny{+-}|10),
\nonumber
\\\nonumber
\BB_2 &=p(\tiny{+-}|00) + p(\tiny{+-}|11) +p(\tiny{++}|01) + p(\tiny{++}|10),\\\nonumber
\BB_{3}&=p(\tiny{+-}|00) + p(\tiny{++}|11) +p(\tiny{+-}|01) + p(\tiny{+-}|10),\\
\BB_4&=p(\tiny{+-}|00) + p(\tiny{+-}|11) + p(\tiny{+-}|01) + p(\tiny{++}|10).\label{bb}
\end{align}
The algebraic maximum of the quantities $\BB_i$, i.e., their maximal possible value,  is  attained at the corresponding extreme point  and is given by $\BB_i=4$. In order to reach $\BB_1=4$ independent of whether $\mathcal{M}_0$ or $\mathcal{M}_1$ is performed, the outcome in the first time step has to be $"+"$. If the same measurement is repeated in the second time step one has to obtain outcome $"+"$ again. Performing a different measurement in the second time step has to yield deterministically $"-"$. This, however, is not possible to accomplish with measurements on a qubit and consequently one can find upper bounds on $\BB_i$ for qubits.
In particular, the following upper bounds have been found   in  Ref. \cite{temporal} for general measurements (see Appendix A  for more details) on a qubit
 \begin{align}\nonumber
\BB_{1}&\stackrel{D=2}{\leq} \mathcal{C}_1 = 3,\,\,\,\,\,\,\,\,\,\,\,\,\,\,\,\,
\BB_{2}\stackrel{D=2}{\leq} \mathcal{C}_2 = 3,
\\
\BB_{3}& \stackrel{D=2}{\leq} \mathcal{C}_3 \approx 3.186,\,\,\,\,\,
\BB_{4} \stackrel{D=2}{\leq} \mathcal{C}_4 \approx 3.186.\label{inequBB}
\end{align}
If the value of $\BB_{i}^{\rm exp}$ obtained in an experiment exceeds the upper bound  this certifies that the system is at least three-dimensional. It is possible to reach the algebraic maximum of $\BB_i=4$  with measurements on a qutrit (see \cite{temporal}, Appendix C and below for corresponding protocols). 

The dimension witnesses based on sequences of length two may still be interpreted as a P\&M scheme where the first (second) measurement corresponds to the state preparation (measurement step) in the P\&M scenario respectively. It is, however, not clear how to interpret the probabilities obtained in the first time step for the sequential witnesses $\BB_i$ in a standard P\&M scenario. To go clearly beyond the P\&M scenario we consider the following dimension witness that makes use of temporal correlations among three time steps 

\begin{align}\nonumber\mathcal{T}&=p(\tiny{+++}|000)+p(\tiny{++-}|001)+p(\tiny{+--}|010)\\\nonumber&+p(\tiny{+-+}|011)+p(\tiny{+-+}|100)+p(\tiny{+--}|101)\\\nonumber&+p(\tiny{++-}|110)+p(\tiny{+++}|111)\\\label{ineq3ts}&\stackrel{D=2}{\leq} \mathcal{C}\approx 5.226.\end{align}
As before, the algebraic maximum given by $\mathcal{T}=8$ is attained for an extreme point of the temporal correlation polytope  which can be obtained with measurements on a qutrit. Hence, a violation of this inequality certifies a dimension of at least three (see Appendix B   for the proof of the upper bound for two-dimensional systems which partially relies on numerics). Moreover, note that the inequality (\ref{ineq3ts}) shows a larger separation between a qubit and a qutrit (in terms of the ratio of their maximal achievable values) than the inequalities presented in Eqs. (\ref{inequBB}).

{\it Experimental setup.---}
In order to obtain the maximal possible violation of the inequalities and to provide an unimpeachable bound on the dimension we prepare the initial state and implement the respective measurements that allow us to reach the corresponding extreme points (for details see below and Appendix C). These can be implemented by using the following building blocks:  a two-outcome projective measurement, the preparation of the system in a pure state and unitaries that permute two levels. In the following we explain the experimental details on the implementation of these building blocks.

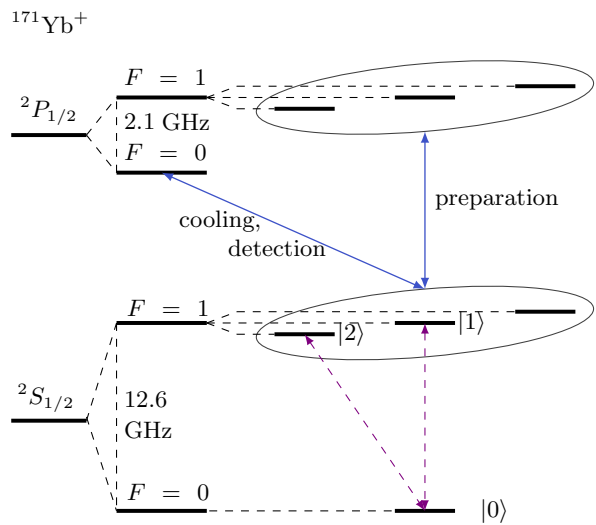
\begin{figure}

\begin{tikzpicture}
\node[anchor=north, text width=2cm](tray1) at (1,6.75){$^{171}\text{Yb}^+$};
\draw[level, >=latex] (0, 1.2) --  (1, 1.2);
\draw[connect, >=latex] (1,1.2) --  (1.4, 2.5);
\draw[connect, >=latex] (1,1.25) --  (1.4, 0);
\draw[connect, color=black] (1.4, 2.5) -- (1.4, 0);
\node[anchor=north,text width=2cm] (tray1) at (2.5,1.7) {$\small{\text{12.6}}$};
\node[anchor=north,text width=2cm] (tray1) at (2.5,1.3) {$\small{\text{GHz}}$};

\draw[level, >=latex] (1.4, 0) --  (2.6, 0);
\node[anchor=east,text width=2cm] (tray1) at (3.65,0.25) {$F=0$};
\node[anchor=east,text width=2cm] (tray1) at (2.2,1.5) {$^{2}S_{1/2}$};
\draw[level, >=latex] (5.1, 0) -- (5.9, 0);
\draw[connect, >=latex] (2,0) --  (5.1, 0);
\node[anchor=east,text width=1cm] (tray1) at (7.35,0) {$\ket{0}$};

\node[anchor=east,text width=2cm] (tray1) at (3.65,2.7) {$F=1$};
\draw[level, >=latex] (1.4, 2.5) -- (2.6, 2.5);

\draw[level, >=latex] (3.5, 2.35) -- (4.3, 2.35);
\node[anchor=east,text width=1cm] (tray1) at (5.45,2.35) {$\ket{2}$};
\draw[level, >=latex] (5.1, 2.5) -- (5.9, 2.5);
\node[anchor=east,text width=1cm] (tray1) at (7.05,2.5) {$\ket{1}$};
\draw[level, >=latex] (6.7, 2.65) -- (7.5, 2.65);
\draw (5.5,2.5) [darkgray] circle [x radius=2.25cm, y radius=4.5mm, rotate=05];

\draw[connect, >=latex] (2.6,2.5) --  (3.0, 2.35);
\draw[connect, >=latex] (3,2.35) --  (3.5, 2.35);

\draw[connect, >=latex] (2.6,2.5) --  (5.1, 2.5);

\draw[connect, >=latex] (2.6,2.5) --  (3, 2.65);
\draw[connect, >=latex] (3,2.65) --  (6.7, 2.65);

\draw[level, >=latex] (0, 5) -- node[above]{$^{2}P_{1/2}$} (1, 5);

\draw[level, >=latex] (1.4, 5.5) -- (2.6, 5.5);
\draw[connect, >=latex] (1,5) --  (1.4, 5.5);
\node[anchor=north,text width=2cm] (tray1) at (2.5,6) {$F=1$};
\draw[level, >=latex] (1.4, 4.5) -- (2.6, 4.5);
\node[anchor=north,text width=2cm] (tray1) at (2.5,5) {$F=0$};
\draw[connect, >=latex] (1, 5) -- (1.4, 4.5);

\draw[connect, >=latex, color=black] (1.4, 4.5) -- (1.4, 5.5);
\node[anchor=north,text width=2cm] (tray1) at (2.5,5.4) {$\small{\text{2.1}}$};
\node[anchor=north,text width=2cm] (tray1) at (3,5.4) {$\small{\text{GHz}}$};

\draw[level, >=latex] (3.5, 5.35) -- (4.3, 5.35);
\draw[level, >=latex] (5.1, 5.5) -- (5.9, 5.5);
\draw[level, >=latex] (6.7, 5.65) -- (7.5, 5.65);
\draw (5.5,5.5) [darkgray] circle [x radius=2.25cm, y radius=4.5mm, rotate=05];

\draw[connect, <->, >=latex, color=lila] (3.9, 2.35) -- (5.5, 0);
\draw[connect, <->, >=latex, color=lila] (5.5, 2.5) -- (5.5, 0);

\draw[connect, >=latex] (2.6,5.5) --  (3, 5.35);
\draw[connect, >=latex] (3,5.35) --  (3.5, 5.35);

\draw[connect, >=latex] (2.6,5.5) --  (5.1, 5.5);

\draw[connect, >=latex] (2.6,5.5) --  (3, 5.65);
\draw[connect, >=latex] (3,5.65) --  (6.7, 5.65);

\draw[transition, <->, >=latex, color=cobalt] (5.5, 2.95) -- (2, 4.5);
\node[anchor=east,text width=1cm] (tray1) at (4.0,3.5) {$\text{detection}$};
\node[anchor=east,text width=1cm] (tray1) at (3.35,3.85) {$\text{cooling,}$};
\draw[transition, <->, >=latex, color=cobalt] (5.5, 2.95) -- (5.5, 5.05);
\node[anchor=east,text width=1cm] (tray1) at (6.75,4.15) {$\text{preparation}$};

\end{tikzpicture}
\caption{Energy levels and transitions relevant for preparation, manipulation and detection of the $^{171}\text{Yb}^+$ ion. The energy splittings are not to scale. A radio frequency (rf) field near 12.6$\,$GHz is used for coherent manipulation of the qutrit. With specific shifts of the rf the transitions $\ket{0} \leftrightarrow \ket{1}$ and $\ket{0} \leftrightarrow \ket{2}$ between the states can be driven. The detection is realized by resonance fluorescence from the ion exited by a laser near 369$\,$nm that drives the transition between $\ket{^2P_{1/2}, F=0}$ and $\ket{^2S_{1/2}, F=1}$. By shifting the cooling/detection laser by 2.1$\,$GHz, the ion can be prepared in $\ket{0}$ by a decay from $\ket{^2P_{1/2}, F=1}$ to $\ket{^2S_{1/2}, F=0}$. The arrows indicate the driven transitions by laser. The dashed arrows indicate the rf driven transitions.}\label{term scheme}\end{figure}

The experiments are performed using a Doppler cooled $^{171}\text{Yb}^+$ ion, stored in a three-layered, micro-structured, segmented Paul trap \cite{Johanning 2011, Kaufmann 2012}. The qutrit is encoded in the Zeeman manifolds of the hyperfine split $^2\text{S}_{1/2}$ electronic ground state, namely $\ket{0} \equiv \ket{F=0, m_F =0}$, $\ket{1} \equiv \ket{F=1, m_F =0}$,  and $\ket{2} \equiv \ket{F=1, m_F =-1}$ (see Fig.~\ref{term scheme}) \cite{Vitanov 2015, Gloger 2015, Kaufmann 2018}.

The preparation of the state $\ket{0}$ is realized by driving the transition between the states $\ket{\text{S}_{1/2}, \text{F}=1}$ and $\ket{\text{P}_{1/2}, \text{F}=1}$ with a 369$\,$nm laser light, followed by spontaneous decay back into the states $\ket{^2S_{1/2}}$. The decay into the state $\ket{^2S_{1/2}, F=0}$ decouples the ion from the driving field as this state does not interact resonantly with the 369$\,$nm laser light. 

The unitaries 
\begin{equation}
\pi_{01} \equiv -i \left( \ket{0}\bra{1}+ \ket{1}\bra{0}\right) + e^{i\varphi_2} \ket{2}\bra{2} \\
\end{equation}
and
\begin{equation}
\pi_{02} \equiv -i \left(\ket{0}\bra{2} + \ket{2}\bra{0} \right) + e^{i\varphi_1} \ket{1}\bra{1}
\end{equation}
are implemented up to a global phase by resonantly driving the transitions $\ket{0}$ $\leftrightarrow$ $\ket{k}$ (k=1,2) using radio frequency (rf) radiation (see Fig.~\ref{term scheme}) \cite{Vitanov 2015, Piltz2016}.

The idling unitary
\begin{equation}
\text{I}\equiv  \ket{0}\bra{0} +e^{i\varphi_1'} \ket{1}\bra{1}+ e^{i\varphi_2'} \ket{2}\bra{2} \end{equation}
is implemented by a free evolution (no driving field). All sequence timings are equalized by adding a pause block to ensure that all pulses and the detection are carried out at the same time respective to the line trigger. Unintended (but fixed) phase shifts are labelled by $\varphi_k$ and $\varphi_k'$.

For state detection, the transition between the states $\ket{^2S_{1/2}, F=1}$ and the state $\ket{^2P_{1/2}, F=0}$ is driven by a near resonant laser at 369$\,$nm. Fluorescence photons generated by spontaneous decay back to the $\ket{^2S_{1/2}}$-states
are detected by an electron multiplying CCD camera. The implementation of a projective measurement is realized by discriminating the non-fluorescing dark state $\ket{0}$ (corresponding to the measurement operator $\ket{0}\bra{0}$) against the indistinguishable bright states $\ket{1}$ and $\ket{2}$ (corresponding to $\ket{1}\bra{1}+\ket{2}\bra{2}$) \cite{Nagourney 1986, Sauter 1986, Bergquist 1986, Vitanov 2015}.  

{\it Implementation of the dimension witnesses.---}
In the experiment the measurement sequences are performed that allow one to reach the extreme points leading to the algebraic maximum of the considered quantities. 
The corresponding pulse sequences are given in Table \ref{measurements}. Let us consider the measurements that lead to the maximum value of $\BB_1$ in more detail. The system is initially prepared in the state $\ket{0}$. The pulse sequences for measurement $\mathcal{M}_0$ realize the projectors $\mathcal{E}_{\tiny{+}|0}= \vert 0 \rangle \langle 0 \vert+ \vert 1 \rangle \langle 1 \vert$ corresponding to outcome $"+"$ and $\mathcal{E}_{\tiny{-}|0}= \vert 2 \rangle \langle 2 \vert$ associated to outcome $"-"$. The post-measurement state is $\ket{1}$ irrespective of the outcome. For the measurement $\mathcal{M}_1$ one obtains projectors of the form  $\mathcal{E}_{\tiny{+}|1}= \vert 0 \rangle \langle 0 \vert+ \vert 2 \rangle \langle 2 \vert$ and $\mathcal{E}_{\tiny{-}|1}= \vert 1 \rangle \langle 1 \vert$. The state after this measurement is $\ket{2}$ for both outcomes. Note that both instruments are implemented by first applying a projective measurement and then preparing the system in a state which only depends on the setting (not on the outcome). It can be easily seen that therefore these measurement allow us to attain the maximal possible value for  $\BB_1$. The measurements leading to the  algebraic maximum of the quantities are explained in Appendix C.
Note  that the measurements that provide the maximal possible violation of the inequality $\mathcal{T}\leq \mathcal{C}$ are the same as the ones for the maximal violation of $\BB_1\leq \mathcal{C}_1$.

The measurements given in Table \ref{measurements} are then combined yielding the full sequence, i.e., the full sequence starts with a preparation P$_0$ of the state $\vert 0 \rangle$  followed by $\mathcal{M}_x$ and $\mathcal{M}_y$ ($x, y \in \lbrace0,1\rbrace$).
After each measurement sequence Doppler cooling 
is performed in order to prepare the ion for the next measurement sequence (see also Appendix C). To achieve a low statistical uncertainty and a significant violation, all sequences are repeated at least 1000 times.

\begin{table}[t]\centering
\begin{tabular}{|c| c |c| c |c|}\hline
 & Meas. & Pulse Sequence & Outcome B&Exp. Values  \\\hline
$\BB_1$ & $\mathcal{M}_0$ & $\pi_{_{02}} \, \text{D} \, \text{C} \, \text{P}_0 \, \pi_{_{01}}$ & $+$  &$3.65\pm 0.06$\\\cline{2-4}
 & $\mathcal{M}_1$ & $\pi_{_{01}} \, \text{D} \, \text{C} \, \text{P}_0 \, \pi_{_{02}}$ & $+$ & \\\hline
$\BB_2$ & $\mathcal{M}_0$ & $\pi_{_{01}} \, \text{D} \, \text{C} \, \text{P}_0 \, \pi_{_{01}}$ & $+$ &$3.66 \pm 0.06$ \\\cline{2-4}
& $\mathcal{M}_1$ & $\pi_{_{02}} \, \text{D} \, \text{C} \, \text{P}_0 \, \pi_{_{02}}$ & $+$  &\\\hline
$\BB_3$ & $\mathcal{M}_0$ & $\text{I}\, \text{D} \, \text{C} \, \text{P}_0 \, \pi_{_{01}}$ & $-$ &$3.75 \pm 0.06$ \\\cline{2-4}
 & $\mathcal{M}_1$ & $\pi_{_{01}} \, \text{D} \, \text{C} \, \text{P}_0 \, \pi_{_{02}}$ & $+$ &\\\hline
$\BB_4$ & $\mathcal{M}_0$ & $\pi_{_{01}} \, \text{D} \, \text{C} \, \text{P}_0 \, \pi_{_{01}}$ & $+$ &$3.70 \pm  0.06$\\\cline{2-4}
 & $\mathcal{M}_1$ & $\text{I} \, \text{D} \, \text{C} \, \text{P}_0 \, \pi_{_{02}}$ & $-$  &\\\hline
$\mathcal{T}$ & $\mathcal{M}_0$ & $\pi_{_{02}} \, \text{D} \, \text{C} \, \text{P}_0 \, \pi_{_{01}}$ & $+$ &$7.00 \pm 0.05$ \\\cline{2-4}
& $\mathcal{M}_1$ & $\pi_{_{01}} \, \text{D} \, \text{C} \, \text{P}_0 \, \pi_{_{02}}$ & $+$ & \\\hline

\end{tabular}
\caption{Pulse sequences for the optimal measurements and experimental values for the quantities $\BB_i$ and $\mathcal{T}$ (cf. Eqs. (\ref{bb}) and (\ref{ineq3ts})). The pulse sequences have to be read from left to right. Each measurement $\mathcal{M}_i$ is started by a sequence of a unitary $\pi_{0\text{k}}$ or idling ($\text{I}$) for the same duration. The pulse is followed by detection (D), Doppler cooling (C), re-preparation in $\vert 0 \rangle$ (P$_0$) and another unitary. It is indicated whether the measurement result $+$ or $-$ is assigned to the observation of fluorescence (B) in the detection step. The values for the  quantities $\BB_{i}^{\rm exp}$ and $\mathcal{T}^{\rm exp}$  obtained from the experimental data are listed.}\label{measurements}
\end{table}

{\it Results.---}
The values obtained in the experiment for the quantities $\BB_{i}^{\rm exp}$ and $\mathcal{T}^{\rm exp}$ are given in Table \ref{measurements}. The observed violations of the temporal inequalities certify a dimension of at least three for the quantum system that is manipulated in the experiment. The error for $\BB_{i}^{\rm exp}$ and $\mathcal{T}^{\rm exp}$ given in Table \ref{measurements} correspond to the confidence intervals of $68$ \% calculated using Hoeffding's tail inequality \cite{hoeffding} (see Appendix E). The algebraic maxima of the quantities are not achieved due to the limited detection fidelities which are given for a bright state by around 0.96 and for a dark state by 0.98 in the experiment using the single threshold method. These fidelities explain the deviation well, as they restrict  the maximal achievable value to approximately 3.8 for the quantities $\BB_{i}^{\rm exp}$ and to around 7.22 for  $\mathcal{T}^{\rm exp}$. 

It should be noted that the temporal inequality based on the sequence of length three does not only show a higher value of violation in absolute terms but also provides a larger violation ratio $\mathcal{T}^{\rm exp}/\mathcal{C}=1.34 \pm 0.01 > \BB_{i}^{\rm exp}/\mathcal{C}_i$ for all $i\in\{1,2,3,4\}$.

{\it Dimension analysis.---}
In order to derive quantitative statements about the dimension of the measured system we consider the following scenario. 
Let us assume that we have access to a qutrit and a qubit and for each of them we can implement the protocol that allows us to obtain the corresponding maximum of $\BB_i$, i.e., $4$ or $\mathcal{C}_i$ respectively. Then we determine the frequency $p_i$ of making use of the qutrit that is necessary to achieve a value of $\BB_{i}^{\rm exp}$, i.e., $4 p_i + (1-p_i) \mathcal{C}_i = \BB_{i}^{\rm exp}$. Hence, we have $p_i=(\BB_{i}^{\rm exp}-\mathcal{C}_i)/(4-\mathcal{C}_i)$ and analogously $q=(\mathcal{T}^{\rm exp}-\mathcal{C})/(8-\mathcal{C})$ with $q$ being the frequency of using the qutrit that is required to obtain the observed value of $\mathcal{T}^{\rm exp}$.   We obtain for the observed data that  
\begin{align}\label{expp0}
q&= 0.64\pm 0.02,\,\,\,\,\,\, p_1=  0.65 \pm 0.06,\\\label{expp1}
p_{2}&= 0.66 \pm 0.06,\,\,\,\,\,\, p_3= 0.70\pm 0.07,\\\label{expp2}
p_4&= 0.64 \pm 0.07.
\end{align}
This shows that for all $\BB_{i}^{\rm exp}$  and $\mathcal{T}^{\rm exp}$ in more than 60~\% of the cases a qutrit has been used.

{\it Critical analysis.---}
Quantum mechanics predicts that the correlations fulfill the AoT constraints. 
For sequences of length two these are given by \begin{equation}\label{eq:cond1AoT}
\sum_b p(ab\vert xy) = \sum_b p(ab\vert xy') \quad \mbox{ for all } a,x, y, y'.
\end{equation}
We critically examined the data used to calculate the quantities $\BB_{i}^{\rm exp}$ and $\mathcal{T}^{\rm exp}$ in the previous section with a likelihood-ratio test to evaluate the probability of obtaining a violation of the AoT constraints that is at least as high as the observed one under the assumption that the AoT constraints hold true, see Appendix E.
We obtain that for our data the  violation of the AoT constraints  are at most equivalent to  $1.4$ standard deviations and hence are statistically not significant.

{\it Conclusion \& Perspectives.---}
We presented a temporal inequality based on measurement sequences of length three whose violation certifies a dimension of at least three. By relying on such longer sequences this inequality does not allow for an interpretation in the P\&M scenario. Moreover, it provides a larger ratio of violation than the dimension witnesses based on sequences of length two that have been introduced previously. We implemented these tests in an experiment usng a single trapped $^{171}$Yb$^+$ ion and we certified the access to a three--dimensional system. Note that in the implementation no superposition of quantum states appears, i.e., the extreme points can be attained with a classical strategy using a three- dimensional system. It would be of relevance to characterize which temporal correlations can be attained with a classical $D$-dimensional system (in comparison to a qudit) and to provide methods distinguishing them.

In our analysis we find that sequences of length three offer stronger tests on the dimensionality than sequences of length two. Hence, it is of interest to investigate whether the violation of the inequalities can increase exponentially with the length of the sequence. Moreover, it would be important to investigate further how to exploit temporal correlations  to identify a benchmark based on state preparation and measurements and to use this to test quantum devices.

We thank Costantino Budroni and Jannik Hoffmann  for helpful discussions.
This work has been supported by  the ERC (Consolidator Grant 683107/TempoQ), the DFG, the Austrian Science Fund (FWF): J 4258-N27, the FQXi Large Grant ``The Observer Observed: A Bayesian Route to the Reconstruction of Quantum Theory", FIS2015-67161-P (MINECO/FEDER) and Basque Government (project IT986-16).

{\it Note added.---}
After completion of this manuscript the dimension witness proposed in \cite{vioquantwiteq}  has been implemented using the IBM quantum experience \cite{expIBM}.

\section{Appendix A: Background}
As described in the main text we examine sequences of general measurements on a quantum state  resulting in correlations $p(abc\ldots|xyz\ldots)$.  In particular, we consider the most general measurements that are allowed within quantum mechanics. Such measurements can be described in terms of instruments \cite{instrument}. These are collections of completely positive  trace-nonincreasing maps $\{\mathcal{I}_{a|x}\}_a$ such that $\sum_a \mathcal{I}_{a|x}$ is trace-preserving. The action of these maps on a quantum state $\rho$ can be written as $\mathcal{I}_{a|x}(\rho)=\sum_{i} \mathcal{M}_i \rho  (\mathcal{M}_i)^\dagger$ with some Kraus operators $\mathcal{M}_i$ that depend on the outcome and the measurement.The probabilities are computed as $p(a|x)=\tr [\mathcal{I}_{a|x}(\vrho_{\rm in} )]$ and the respective (unnormalized) post-measurement states are given by \begin{equation}
\varrho_{\rm out} = \II_{a|x}(\varrho_{\rm in}).
\end{equation}
Any instrument can be implemented by coupling the system with a unitary transformation (that depends on the instrument) to an ancilla and performing a projective measurement on the ancilla  (see e.g. \cite{instrument}).
In case, one is interested in the probability of a single measurement it is sufficient to consider the effects $\EE_{a|x}=\sum_{i} (\mathcal{M}_i)^\dagger\mathcal{M}_i$ defined by the instrument $\II_{a|x}$ due to $p(a|x)=\tr [\EE_{a|x}\vrho_{\rm in} ]$. Note that if there is no time evolution between the measurements then the states that are measured within the sequence can be related. This is due to the fact that they are obtained by successive application of the corresponding maps $\mathcal{I}_{a|x}$  on the initial state.
Recall that we allow for arbitrary transformations between the measurements which can depend on previous outcomes, measurements and on the time step. This implies that in the scenario we consider here the effects of a measurement have to stay the same at each time step, however, the state that are measured within the sequence do not need to be related anymore but are completely arbitrary.

\section{Appendix B: Dimension witness based on a sequence of length 3}

In the following we consider the quantity $\mathcal{T}=p(\tiny{+++}|000)+p(\tiny{++-}|001)+p(\tiny{+--}|010)+p(\tiny{+-+}|011)+p(\tiny{+-+}|100)+p(\tiny{+--}|101)+p(\tiny{++-}|110)+p(\tiny{+++}|111)$ constructed from temporal sequences of length three and investigate its maximum for measurements on a qubit. In particular, we prove that it holds that $\mathcal{T}\stackrel{D=2}{\leq} 5.226$. The proof is analogous to the one used to derive the bounds for two-dimensional systems on the quantities $\BB_i$ in \cite{temporal}. Note that this raises hope that one may be able to show dimension bounds also for quantities based on even longer sequences.

In order to prove the bound on $\mathcal{T}$ we first show that either the maximum of  $\mathcal{T}$ is attained if  for both measurements the effect for outcome $"-"$ is proportional to a projector [case (i)] or  $\mathcal{T}\leq 5$ [case (ii)]. For case (i) we identify then  the optimal initial and post-measurement states and perform a numerical optimization (over 3 parameters) that strongly suggests that the maximal value of $\mathcal{T}$ that can be attained with measurements on a qubit is given by $5.226$ and is reached if the effects are projective for both measurements.\\
Let us first define our notation. The effects for the measurement setting $s$  and outcome $r\in \{\tiny{+},\tiny{-}\}$ will be denoted by $\EE_{r|s}$. First the following decomposition for these effects will be used,
\begin{align}\label{effapp210}
&\EE_{\tiny{+}|0}=a_0(\openone + b_0\,\vec{c}\cdot\vec{\sigma}),\\\label{effapp220}
&\EE_{\tiny{-}|0}=\openone-\EE_{\tiny{+}|0},\\
&\EE_{\tiny{+}|1}=a_1(\openone + b_1\,\vec{d}\cdot\vec{\sigma}),\\\label{effapp240}
&\EE_{\tiny{-}|1}=\openone-\EE_{\tiny{+}|1},
\end{align}
with $\vec{c},\vec{d}\in\mathbb{R}^3$, $|\vec{c}|=|\vec{d}|=1$ and $\vec{\sigma}=(\sigma_1,\sigma_2,\sigma_3)$ where here $\sigma_i$ denote the Pauli matrices. Without loss of generality we choose that $b_s\geq 0$ and hence due to $\openone\geq \EE_{r|s}\geq 0$ it has to hold that  $0\leq a_s\leq \frac{1}{1+b_s}$ and $b_s\leq 1$ for $s\in\{0,1\}$. Moreover, we denote by $\vrho_{\rm in}$ the initial state and by $\vrho_{x}$ the post-measurement states given that measurement $\mathcal{M}_x$ has been applied at the first time step  and by $\vrho_{xy}$ the post-measurement states given that in the first time step measurement $x$ and at the second time step measurement $y$ has been implemented (with the respective outcomes as they occur in $\mathcal{T}$). Note that by post-measurement state we refer to the state that is measured at the successive time step, i.e., the state after the measurement potentially followed by some state transformation. We parametrize these states via their Bloch decomposition \bea\label{eqrho0}
\vrho_j=\frac{1}{2}(\openone+\vec{\alpha}_{j}\cdot\vec{\sigma})
\eea
for $j\in\{{\rm in}, 0, 1, 00, 01, 10, 11\}$ where $\vec{\alpha}_{j}\in\mathbb{R}^3$ and $|\vec{\alpha}_{j}|\leq 1$. 
It has been observed that if $p(abc|xyz)$ fulfills the AoT constraints it can be written as $p(abc|xyz)=p(a|x) p(b|axy) p(c|abxyz)$ \cite{temporal, Fritz}. Here $p(b|axy)$ is the probability for obtaining outcome $b$ when performing measurement $y$ in the second time step given that in the first time step measurement $x$ has been implemented and outcome $a$ has been observed. Analogously,  $p(c|abxyz)$ is the conditional probability distribution obtained for the third time step given the corresponding previous outcomes and measurements. Using that the correlations factorize one obtains that
\begin{align}\nonumber\mathcal{T}&=p(\tiny{+}|0)\big\{p(\tiny{+}|\tiny{+}00)\big[p(\tiny{+}|\tiny{++}000)+p(\tiny{-}|\tiny{++}001)\big]\\\nonumber&+p(\tiny{-}|\tiny{+}01)\big[p(\tiny{-}|\tiny{+-}010)+p(\tiny{+}|\tiny{+-}011)\big]\big\}\\\nonumber&
+p(\tiny{+}|1)\big\{p(\tiny{-}|\tiny{+}10)\big[p(\tiny{+}|\tiny{+-}100)+p(\tiny{-}|\tiny{+-}101)\big]\\\label{eq0BB}&+p(\tiny{+}|\tiny{+}11)\big[p(\tiny{-}|\tiny{++}110)+p(\tiny{+}|\tiny{++}111)\big]\big\}.
\end{align}
We will next show that for arbitrary initial and post-measurement states $\mathcal{T}\leq 5$  or its maximum is attained if for both measurements the effects corresponding to outcome $"-"$ have rank 1.
In order to do so we consider the decomposition of effects given in Eqs. (\ref{effapp210}) and (\ref{effapp220}) and calculate the points where the derivative of Eq. (\ref{eq0BB}) with respect to $a_0$ vanishes (all other parameters are assumed to be fixed, but arbitrary). We obtain that 
\begin{align}\nonumber
a_0\frac{d\mathcal{T}}{d a_0}&= p(\tiny{+}|0)\{p(\tiny{+}|\tiny{+}00)[p(\tiny{+}|\tiny{++}000)+p(\tiny{-}|\tiny{++}001)]\\\nonumber&+p(\tiny{-}|\tiny{+}01)[p(\tiny{-}|\tiny{+-}010)+p(\tiny{+}|\tiny{+-}011)]\}\\\nonumber&+p(\tiny{+}|0)p(\tiny{+}|\tiny{+}00) [p(\tiny{+}|\tiny{++}000)+p(\tiny{-}|\tiny{++}001)]\\\nonumber&+ p(\tiny{+}|0)p(\tiny{+}|\tiny{+}00)p(\tiny{+}|\tiny{++}000)\\\nonumber&+p(\tiny{+}|0)p(\tiny{-}|\tiny{+}01)[p(\tiny{-}|\tiny{+-}010)-1]\\\nonumber&+p(\tiny{+}|1)p(\tiny{-}|\tiny{+}10)p(\tiny{+}|\tiny{+-}100)\\\nonumber&+p(\tiny{+}|1)p(\tiny{+}|\tiny{+}11)[p(\tiny{-}|\tiny{++}110)-1]\\\nonumber&+p(\tiny{+}|1)[p(\tiny{-}|\tiny{+}10)-1][p(\tiny{+}|\tiny{+-}100)+p(\tiny{-}|\tiny{+-}101)]\\\nonumber
&=\mathcal{T}+p(\tiny{+}|0)p(\tiny{+}|\tiny{+}00) [p(\tiny{+}|\tiny{++}000)+p(\tiny{-}|\tiny{++}001)]\\\nonumber&+ p(\tiny{+}|0)p(\tiny{+}|\tiny{+}00)p(\tiny{+}|\tiny{++}000)\\\nonumber&+p(\tiny{+}|0)p(\tiny{-}|\tiny{+}01)p(\tiny{-}|\tiny{+-}010)\\\nonumber&+p(\tiny{+}|1)p(\tiny{-}|\tiny{+}10)p(\tiny{+}|\tiny{+-}100)\\\nonumber&-p(\tiny{+}|1)p(\tiny{+}|\tiny{+}11)-p(\tiny{+}|1)p(\tiny{+}|\tiny{+}11)p(\tiny{+}|\tiny{++}111)\\\nonumber&-p(\tiny{+}|1)[p(\tiny{+}|\tiny{+-}100)+p(\tiny{-}|\tiny{+-}101)]\\\nonumber&
-p(\tiny{+}|0)p(\tiny{-}|\tiny{+}01)\stackrel{!}{=}0.
\end{align}
This implies that for the points where the derivative vanishes
\begin{align}\nonumber
\mathcal{T}&=p(\tiny{+}|1)p(\tiny{+}|\tiny{+}11)+p(\tiny{+}|1)p(\tiny{+}|\tiny{+}11)p(\tiny{+}|\tiny{++}111)\\\nonumber&+p(\tiny{+}|1)[p(\tiny{+}|\tiny{+-}100)+p(\tiny{-}|\tiny{+-}101)]\\\nonumber&
+p(\tiny{+}|0)p(\tiny{-}|\tiny{+}01)-p(\tiny{+}|1)p(\tiny{-}|\tiny{+}10)p(\tiny{+}|\tiny{+-}100)\\\nonumber&
-p(\tiny{+}|0)p(\tiny{+}|\tiny{+}00) [p(\tiny{+}|\tiny{++}000)+p(\tiny{-}|\tiny{++}001)]\\\nonumber&- p(\tiny{+}|0)p(\tiny{+}|\tiny{+}00)p(\tiny{+}|\tiny{++}000)\\&-p(\tiny{+}|0)p(\tiny{-}|\tiny{+}01)p(\tiny{-}|\tiny{+-}010)\leq 5.
\end{align}
 It remains then to consider also the values of $\mathcal{T}$ at the boundary  of the domain $a_0\in [0, 1/(1+b_0)]$. For $a_0=0$ it holds that $\mathcal{T}\leq 3$. Hence, either $\mathcal{T}\leq 5$ or it attains its maximum value at $a_0= 1/(1+b_0)$. Note that $\mathcal{T}$ is symmetric under the exchange of the two measurements and therefore it holds analogously that the maximum of $\mathcal{T}$ is smaller or equal $5$ or it is obtained for $a_1= 1/(1+b_1)$. Considering the case $a_s= 1/(1+b_s)$ for $s=0, 1$ and substituting $b_0=p/(2-p)$ and $b_1=q/(2-q)$ the effects can then be written as 
\begin{align}\label{effapp10}
&\EE_{\tiny{+}|0}=\frac{1}{2}[(2-p) \openone + p \,\vec{c}\cdot\vec{\sigma}],\\\label{effapp400}
&\EE_{\tiny{-}|0}=\frac{p}{2}(\openone -\, \vec{c}\cdot\vec{\sigma}),\\\label{effapp401}
&\EE_{\tiny{+}|1}=\frac{1}{2}[(2-q) \openone + q \,\vec{d}\cdot\vec{\sigma}],\\\label{effapp402}
&\EE_{\tiny{-}|1}=\frac{q}{2}(\openone -\, \vec{d}\cdot\vec{\sigma}),
\end{align} where $0\leq p\leq 1$ and $0\leq q\leq 1$. 
Inserting this decomposition for $p(c|abxyz)$ and the Bloch decomposition for the states $\vrho_{xy}$ in $\mathcal{T}$ one obtains that 
\begin{align}\nonumber\mathcal{T}&=\frac{p(\tiny{+}|0)}{2}\{p(\tiny{+}|\tiny{+}00)[2-p+q +\vec{\alpha}_{00}\cdot ( p \,\vec{c}-q \,\vec{d})]\\\nonumber&+p(\tiny{-}|\tiny{+}01)[p+2-q -\vec{\alpha}_{01}\cdot ( p \,\vec{c}-q \,\vec{d})]\}\\\nonumber&
+\frac{p(\tiny{+}|1)}{2}\{p(\tiny{-}|\tiny{+}10)[2-p+q +\vec{\alpha}_{10}\cdot ( p \,\vec{c}-q \,\vec{d})]\\&+p(\tiny{+}|\tiny{+}11))[p+2-q -\vec{\alpha}_{11}\cdot ( p \,\vec{c}-q \,\vec{d})]\}.\label{eq1BB}
\end{align}
From Eq. (\ref{eq1BB}) it can be easily seen that it is optimal to choose
\bea\nonumber
&\vec{\alpha}_{00}=\vec{\alpha}_{10}= -\vec{\alpha}_{01}=-\vec{\alpha}_{11}=\frac{p\vec{c}-q\vec{d}}{\sqrt{p^2+q^2-2pq\cos (\gamma )}}\,\,\\& \mathrm{if} \,\,p\vec{c}\neq q\vec{d},\eea  
where $\cos (\gamma)=\vec{c}\cdot\vec{d}$.  Note that if $p\vec{c}=q\vec{d}$ then  Eq. (\ref{eq1BB}) does not depend on $\vec{\alpha}_{j}$ for $j\in\{00, 01, 10, 11\}$.
With this and the notation $X_0=2-p+q +\sqrt{p^2+q^2-2pq\cos (\gamma )}$ and $X_1=p+2-q +\sqrt{p^2+q^2-2pq\cos (\gamma )}$ we obtain
\begin{align}
&\mathcal{T}=\frac{p(\tiny{+}|0)}{2}[p(\tiny{+}|\tiny{+}00)X_0+p(\tiny{-}|\tiny{+}01)X_1]\\\nonumber&
+\frac{p(\tiny{+}|1)}{2}[p(\tiny{-}|\tiny{+}10)X_0+p(\tiny{+}|\tiny{+}11)X_1]\\\nonumber&=\frac{p(\tiny{+}|0)}{4}[(2-p)X_0+q X_1+\vec{\alpha}_{0}\cdot (p X_0\vec{c}-qX_1\vec{d})]\\\nonumber&+\frac{p(\tiny{+}|1)}{4}[pX_0+(2-q)X_1-\vec{\alpha}_{1}\cdot (p X_0\vec{c}-qX_1\vec{d})].
\end{align}
This quantity is maximized by choosing
\bea\nonumber
&\vec{\alpha}_{0}=-\vec{\alpha}_{1}=\frac{pX_0\vec{c}-qX_1\vec{d}}{\sqrt{p^2X_0^2+q^2X_1^2-2pqX_0X_1\cos (\gamma )}}\,\, \\&\mathrm{if} \,\,pX_0\vec{c}\neq qX_1\vec{d}.\eea  
Using this as well as the notation $Y_0=(2-p)X_0+q X_1+\sqrt{p^2X_0^2+q^2X_1^2-2pqX_0X_1\cos (\gamma )}$ and $Y_1=pX_0+(2-q)X_1+\sqrt{p^2X_0^2+q^2X_1^2-2pqX_0X_1\cos (\gamma )}$ leads to 
\begin{align}
&\frac{1}{4}[p(\tiny{+}|0)Y_0+p(\tiny{+}|1)Y_1]\\\nonumber&=\frac{1}{8}[(2-p)Y_0+(2-q) Y_1+\vec{\alpha}_{\rm in}\cdot (p Y_0\vec{c}+qY_1\vec{d})],\end{align}
which attains its largest value for 
\bea\nonumber
&\vec{\alpha}_{\rm in}=\frac{pY_0\vec{c}+qY_1\vec{d}}{\sqrt{p^2Y_0^2+q^2Y_1^2+2pqY_0Y_1\cos (\gamma )}}\,\, \\&\mathrm{if} \,\,pY_0\vec{c}\neq -qY_1\vec{d}.\eea  
Hence, for the optimal states and the measurements of the form given in Eqs. (\ref{effapp10})-(\ref{effapp402}) we have that 
\begin{align}
\mathcal{T}&= \frac{1}{8}[(2-p)Y_0+(2-q) Y_1\\\nonumber&+\sqrt{p^2Y_0^2+q^2Y_1^2+2pqY_0Y_1\cos (\gamma )}].
\end{align}
Performing a numerical optimization of the upper bound over the remaining 3 parameters one obtains that $p=q=1$ and $\cos (\gamma )=-0.458$ which yields  $\mathcal{T}\stackrel{D=2}{\leq} 5.226$. For this choice of parameters $\vec{\alpha}_0=\vec{\alpha}_{00}=\vec{\alpha}_{10}$ and $\vec{\alpha}_1=\vec{\alpha}_{11}=\vec{\alpha}_{01}$ which implies that there exist measurements constituting of first applying a projective measurement and then preparing a fixed state  that allow one to obtain the maximum value of $\mathcal{T}$ on a qubit (for a trivial time evolution between the measurements). Moreover, note that  we only imposed that the effects of the measurements are the same at each time step and allowed for arbitrary post-measurement states. This implies that for arbitrary state transformations between the measurements this bound cannot be exceeded with a qubit. $\hfill \Box$

\section{Appendix C: Measurements that lead to maximal violations of the temporal inequalities}

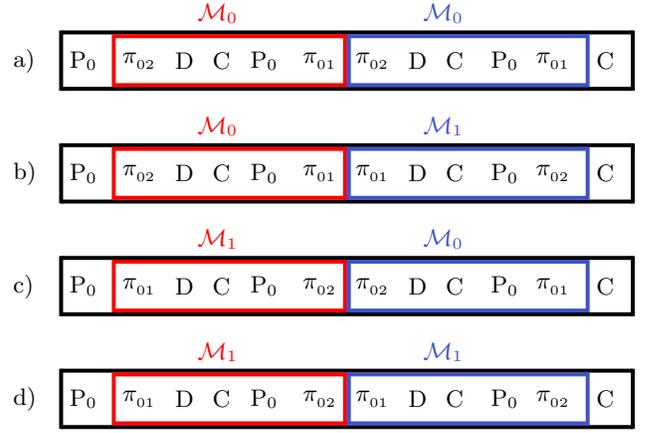
\begin{figure}[t!]

\begin{tikzpicture}


\draw[level, >=latex] (0, 4.5) --  (7.6, 4.5);
\draw[level, >=latex] (0, 5.25) --  (7.6, 5.25);
\draw[level, >=latex, color=red] (0.7, 4.55) --  (3.8, 4.55);
\draw[level, >=latex, color=red] (0.7, 5.2) --  (3.8, 5.2);
\draw[level, >=latex, color=cobalt] (3.8, 4.55) --  (7, 4.55);
\draw[level, >=latex, color=cobalt] (3.8, 5.2) --  (7, 5.2);
\draw[level, >=latex] (0, 4.475) --  (0, 5.275);
\draw[level, >=latex] (7.6, 4.475) --  (7.6, 5.275);
\draw[level, >=latex, color=red] (0.7, 4.525) --  (0.7, 5.225);
\draw[level, >=latex, color=red] (3.775, 4.525) --  (3.775, 5.225);
\draw[level, >=latex, color=cobalt] (3.825, 4.525) --  (3.825, 5.225);
\draw[level, >=latex, color=cobalt] (7, 4.525) --  (7, 5.225);

\node[anchor=west,text width=2cm] (tray1) at (-0.75,4.875) {$\text{a})$};

\node[anchor=west,text width=2cm] (tray1) at (0,4.875) {$\text{P}_0$};
\node[anchor=west,text width=2cm] (tray1) at (0.7,4.875) {$\pi_{_{02}}$};
\node[anchor=west,text width=2cm] (tray1) at (1.4,4.875) {$\text{D}$};
\node[anchor=west,text width=2cm] (tray1) at (1.9,4.875) {$\text{C}$};

\node[anchor=west,text width=2cm] (tray1) at (2.4,4.875) {$\text{P}_0$};
\node[anchor=west,text width=2cm] (tray1) at (3.1,4.875) {$\pi_{_{01}}$};
\node[anchor=west,text width=2cm] (tray1) at (3.8,4.875) {$\pi_{_{02}}$};

\node[anchor=west,text width=2cm] (tray1) at (4.5,4.875) {$\text{D}$};
\node[anchor=west,text width=2cm] (tray1) at (5.0,4.875) {$\text{C}$};

\node[anchor=west,text width=2cm] (tray1) at (5.6,4.875) {$\text{P}_0$};
\node[anchor=west,text width=2cm] (tray1) at (6.2,4.875) {$\pi_{_{01}}$};
\node[anchor=west,text width=2cm] (tray1) at (7.0, 4.875) {$\text{C}$};

\node[anchor=west,text width=2cm, color=red] (tray1) at (1.7,5.5) {$\mathcal{M}_0$};
\node[anchor=west,text width=2cm, color=cobalt] (tray1) at (4.7,5.5) {$\mathcal{M}_0$};


\draw[level, >=latex] (0, 3) --  (7.6, 3);
\draw[level, >=latex] (0, 3.75) --  (7.6, 3.75);
\draw[level, >=latex, color=red] (0.7, 3.05) --  (3.8, 3.05);
\draw[level, >=latex, color=red] (0.7, 3.7) --  (3.8, 3.7);
\draw[level, >=latex, color=cobalt] (3.8, 3.05) --  (7, 3.05);
\draw[level, >=latex, color=cobalt] (3.8, 3.7) --  (7, 3.7);
\draw[level, >=latex] (0, 2.975) --  (0, 3.775);
\draw[level, >=latex] (7.6, 2.975) --  (7.6, 3.775);
\draw[level, >=latex, color=red] (0.7, 3.025) --  (0.7, 3.725);
\draw[level, >=latex, color=red] (3.775, 3.025) --  (3.775, 3.725);
\draw[level, >=latex, color=cobalt] (3.825, 3.025) --  (3.825, 3.725);
\draw[level, >=latex, color=cobalt] (7, 3.025) --  (7, 3.725);


\node[anchor=west,text width=2cm] (tray1) at (-0.75,3.375) {$\text{b})$};

\node[anchor=west,text width=2cm] (tray1) at (0, 3.375) {$\text{P}_0$};
\node[anchor=west,text width=2cm] (tray1) at (0.7, 3.375) {$\pi_{_{02}}$};
\node[anchor=west,text width=2cm] (tray1) at (1.4,3.375) {$\text{D}$};
\node[anchor=west,text width=2cm] (tray1) at (1.9,3.375) {$\text{C}$};

\node[anchor=west,text width=2cm] (tray1) at (2.4,3.375) {$\text{P}_0$};
\node[anchor=west,text width=2cm] (tray1) at (3.1,3.375) {$\pi_{_{01}}$};
\node[anchor=west,text width=2cm] (tray1) at (3.8,3.375) {$\pi_{_{01}}$};

\node[anchor=west,text width=2cm] (tray1) at (4.5,3.375) {$\text{D}$};
\node[anchor=west,text width=2cm] (tray1) at (5.0,3.375) {$\text{C}$};

\node[anchor=west,text width=2cm] (tray1) at (5.6,3.375) {$\text{P}_0$};
\node[anchor=west,text width=2cm] (tray1) at (6.2,3.375) {$\pi_{_{02}}$};
\node[anchor=west,text width=2cm] (tray1) at (7.0,3.375) {$\text{C}$};

\node[anchor=west,text width=2cm, color=red] (tray1) at (1.7,4) {$\mathcal{M}_0$};
\node[anchor=west,text width=2cm, color=cobalt] (tray1) at (4.7,4) {$\mathcal{M}_1$};


\draw[level, >=latex] (0, 1.5) --  (7.6, 1.5);
\draw[level, >=latex] (0, 2.25) --  (7.6, 2.25);
\draw[level, >=latex, color=red] (0.7, 1.55) --  (3.8, 1.55);
\draw[level, >=latex, color=red] (0.7, 2.2) --  (3.8, 2.2);
\draw[level, >=latex, color=cobalt] (3.8, 1.55) --  (7, 1.55);
\draw[level, >=latex, color=cobalt] (3.8, 2.2) --  (7, 2.2);
\draw[level, >=latex] (0, 1.475) --  (0, 2.275);
\draw[level, >=latex] (7.6, 1.475) --  (7.6, 2.275);
\draw[level, >=latex, color=red] (0.7, 1.525) --  (0.7, 2.225);
\draw[level, >=latex, color=red] (3.775, 1.525) --  (3.775, 2.225);
\draw[level, >=latex, color=cobalt] (3.825, 1.525) --  (3.825, 2.225);
\draw[level, >=latex, color=cobalt] (7, 1.525) --  (7, 2.225);


\node[anchor=west,text width=2cm] (tray1) at (-0.75,1.875) {$\text{c})$};

\node[anchor=west,text width=2cm] (tray1) at (0,1.875) {$\text{P}_0$};
\node[anchor=west,text width=2cm] (tray1) at (0.7,1.875) {$\pi_{_{01}}$};
\node[anchor=west,text width=2cm] (tray1) at (1.4,1.875) {$\text{D}$};
\node[anchor=west,text width=2cm] (tray1) at (1.9,1.875) {$\text{C}$};

\node[anchor=west,text width=2cm] (tray1) at (2.4,1.875) {$\text{P}_0$};
\node[anchor=west,text width=2cm] (tray1) at (3.1,1.875) {$\pi_{_{02}}$};
\node[anchor=west,text width=2cm] (tray1) at (3.8,1.875) {$\pi_{_{02}}$};

\node[anchor=west,text width=2cm] (tray1) at (4.5,1.875) {$\text{D}$};
\node[anchor=west,text width=2cm] (tray1) at (5.0,1.875) {$\text{C}$};

\node[anchor=west,text width=2cm] (tray1) at (5.6,1.875) {$\text{P}_0$};
\node[anchor=west,text width=2cm] (tray1) at (6.2,1.875) {$\pi_{_{01}}$};
\node[anchor=west,text width=2cm] (tray1) at (7.0,1.875) {$\text{C}$};

\node[anchor=west,text width=2cm, color=red] (tray1) at (1.7,2.5) {$\mathcal{M}_1$};
\node[anchor=west,text width=2cm, color=cobalt] (tray1) at (4.7,2.5) {$\mathcal{M}_0$};


\draw[level, >=latex] (0, 0) --  (7.6, 0);
\draw[level, >=latex] (0, 0.75) --  (7.6, 0.75);
\draw[level, >=latex, color=red] (0.7, 0.05) --  (3.8, 0.05);
\draw[level, >=latex, color=red] (0.7, 0.7) --  (3.8, 0.7);
\draw[level, >=latex, color=cobalt] (3.8, 0.05) --  (7, 0.05);
\draw[level, >=latex, color=cobalt] (3.8, 0.7) --  (7, 0.7);
\draw[level, >=latex] (0, -0.025) --  (0, 0.775);
\draw[level, >=latex] (7.6, -0.025) --  (7.6, 0.775);
\draw[level, >=latex, color=red] (0.7, 0.025) --  (0.7, 0.725);
\draw[level, >=latex, color=red] (3.775, 0.025) --  (3.775, 0.725);
\draw[level, >=latex, color=cobalt] (3.825, 0.025) --  (3.825, 0.725);
\draw[level, >=latex, color=cobalt] (7, 0.025) --  (7, 0.725);


\node[anchor=west,text width=2cm] (tray1) at (-0.75,0.375) {$\text{d})$};

\node[anchor=west,text width=2cm] (tray1) at (0,0.375) {$\text{P}_0$};
\node[anchor=west,text width=2cm] (tray1) at (0.7,0.375) {$\pi_{_{01}}$};
\node[anchor=west,text width=2cm] (tray1) at (1.4,0.375) {$\text{D}$};
\node[anchor=west,text width=2cm] (tray1) at (1.9,0.375) {$\text{C}$};

\node[anchor=west,text width=2cm] (tray1) at (2.4,0.375) {$\text{P}_0$};
\node[anchor=west,text width=2cm] (tray1) at (3.1,0.375) {$\pi_{_{02}}$};
\node[anchor=west,text width=2cm] (tray1) at (3.8,0.375) {$\pi_{_{01}}$};

\node[anchor=west,text width=2cm] (tray1) at (4.5,0.375) {$\text{D}$};
\node[anchor=west,text width=2cm] (tray1) at (5.0,0.375) {$\text{C}$};

\node[anchor=west,text width=2cm] (tray1) at (5.6,0.375) {$\text{P}_0$};
\node[anchor=west,text width=2cm] (tray1) at (6.2,0.375) {$\pi_{_{02}}$};
\node[anchor=west,text width=2cm] (tray1) at (7.0,0.375) {$\text{C}$};

\node[anchor=west,text width=2cm, color=red] (tray1) at (1.7,1) {$\mathcal{M}_1$};
\node[anchor=west,text width=2cm, color=cobalt] (tray1) at (4.7,1) {$\mathcal{M}_1$};


\end{tikzpicture}
\caption{Scheme for the complete sequence with the measurements $\mathcal{M}_x$ followed by $\mathcal{M}_y$ for the quantity $\BB_1$. Each measurement includes two $\pi_{0k}$-pulses. Before the measurements starts, the ion is prepared in $\ket{0}$ by optical pumping (P$_0$). After each detection (D) the ion is Doppler cooled (C) and re-prepared by P$_0$. After the last measurement, the generated post measurement state is Doppler cooled in order to reset the qutrit before performing the next measurement.}\label{fullSequence}

\end{figure}

In Fig. \ref{fullSequence} the measurement sequences using the optimal measurements for the quantity $\BB_1$ are depicted in detail. We will next consider the measurements that allow to reach the maximal possible values for the quantities $\BB_i$ for $i=2, 3, 4$  whose pulse sequences are given in Table \ref{measurements} in the main text. The measurements for $\BB_1$ and $\mathcal{T}$ which coincide have been explained in the main text.

Here and in the following the initial state will be $\ket{0}$. The post-measurement state for $\mathcal{M}_0$ is for all quantities and outcomes $\ket{1}$ and the one for $\mathcal{M}_1$ is always $\ket{2}$. More precisely, the instruments are implemented by performing a projective measurement (which one is depending on the quantity $\BB_i$ and the setting) and then preparing the system in a fixed state (which one depends solely on the setting). In order to reach the algebraic maximum of $\BB_2$ the measurement $\mathcal{M}_0$ has the effect  $\mathcal{E}_{\tiny{+}|0}= \vert 0 \rangle \langle 0 \vert+ \vert 2 \rangle \langle 2 \vert$ associated to outcome $"+"$ and $\mathcal{E}_{\tiny{-}|0}= \vert 1 \rangle \langle 1 \vert$ corresponding to outcome $"-"$.  For $\mathcal{M}_1$ the effects are given by $\mathcal{E}_{\tiny{+}|1}= \vert 0 \rangle \langle 0 \vert+ \vert 1 \rangle \langle 1 \vert$ and $\mathcal{E}_{\tiny{-}|1}= \vert 2 \rangle \langle 2 \vert$. It follows straightforwardly that these measurements lead to  $\BB_2=4$.

In order to see that the pulse sequences given in Table \ref{measurements} allow to reach the the algebraic maximum of $\BB_3$ consider first measurement $\mathcal{M}_0$. The effects that are realized with this pulse sequence are given by $\mathcal{E}_{\tiny{+}|0}= \vert 0 \rangle \langle 0 \vert$ and $\mathcal{E}_{\tiny{-}|0}= \vert 1 \rangle \langle 1 \vert+ \vert 2 \rangle \langle 2 \vert$. For the measurement $\mathcal{M}_1$ the effects are  $\mathcal{E}_{\tiny{+}|1}= \vert 0 \rangle \langle 0 \vert+ \vert 2 \rangle \langle 2 \vert$ and $\mathcal{E}_{\tiny{-}|1}= \vert 1 \rangle \langle 1 \vert$. It can be easily seen that with these measurements one obtains the maximal possible value of $\BB_3$.

For the quantity $\BB_4$ the effects for $\mathcal{M}_0$ are given by $\mathcal{E}_{\tiny{+}|0}= \vert 0 \rangle \langle 0 \vert+ \vert 2 \rangle \langle 2 \vert$  and $\mathcal{E}_{\tiny{-}|0}= \vert 1 \rangle \langle 1 \vert$. The ones for $\mathcal{M}_1$ correspond to $\mathcal{E}_{\tiny{+}|1}= \vert 0 \rangle \langle 0 \vert$ and $\mathcal{E}_{\tiny{-}|1}= \vert 1 \rangle \langle 1 \vert+ \vert 2 \rangle \langle 2 \vert$. As can be easily seen these measurements allow to attain the algebraic maximum of $\BB_4$.

It may happen that during the Doppler cooling (see Fig. \ref{fullSequence} and main text)  the ion decays to a metastable state (with low probability). This metastable state shows a low fluorescence rate and is therefore mainly detected as a dark state. Due to this process the measurements that are implemented in the experiment deviate from the ideal measurements explained above.

\section{Appendix D: Results with validation of the sequences}
As mentioned in Appendix C it may happen that during the Doppler cooling the ion decays to a metastable state. Here the monitored cooling fluorescence gives an indicator whether the ion is in the ground state (high fluorescence rate) or in a metastable state (low fluorescence rate). Hence, the fluorescence rate can be used for a validation procedure. If the ion is in a metastable state the subsequent measurement does not start with a properly initialized qutrit and is therefore discarded in  the validation. In Table \ref{validation} the experimental values $\BB_{i}^{\rm exp}$ and $\mathcal{T}^{\rm exp}$  obtained from the experimental data with validation are presented. 

\begin{table}[H]
\centering
\begin{tabular}{|c| c|}\hline
 &Exp. Values  \\\hline
$\BB_1$ &$3.67\pm 0.06$\\ \hline
$\BB_2$ &$3.68 \pm 0.06$ \\\hline
$\BB_3$ &$3.78 \pm 0.06$ \\\hline
$\BB_4$ &$3.74 \pm  0.06$\\\hline
$\mathcal{T}$ & $7.06 \pm 0.05$ \\\hline
\end{tabular}
\caption{Experimental values for the quantities $\BB_i$ and $\mathcal{T}$ with validation of the sequences.}\label{validation}
\end{table}

\section{Appendix E: Confidence intervals and analysis of the AoT constraints}
In this Appendix we first explain the derivation of the confidence intervals for the experimental values of $\BB_{i}^{\rm exp}$ in more detail (see also \cite{Tobias, Schwemmer}).  For $\mathcal{T}^{\rm exp}$ the derivation is analogous. Then we will outline the analysis of the AoT constraints. The derivation of the confidence interval is based on Hoeffding's inequality \cite{hoeffding} which states the following. 

Let $X_i$  be independent random variables with $a_i\leq X_i\leq  b_i$ for $i\in\{1,\ldots, n\}$ and no further assumption on their distribution. Moreover, denote by $\bar{X}=\sum_{i=1}^nX_i/n$ their sample mean and by $\mathbb{E}(\bar{X})$ the mean value of $\bar{X}$.  Then it holds for all $t>0$ that 
\begin{equation}\label{Hoeffd2}
{\rm Prob}[ |\bar{X}-\mathbb{E}(\bar{X})|\geq t]\leq 2\,{\rm exp}\left[-\frac{2n^2t^2}{\sum_i(b_i-a_i)^2}\right].\end{equation}
The random variable $X_i\equiv Z_{(x,y)}^j$ will refer in the following to the $j$-th conduct of the measurement $x$ in the first time step followed by measuring setting $y$ in the second time step. More precisely, depending on the quantity, $\sum_{a,b,x,y} \alpha_{(x,y)}^{(a,b)} p(ab|xy)$, we assign to $Z_{(x,y)}^j$ the value $\alpha_{(x,y)}^{(a,b)} n/n_{(x,y)}$, where  $a$ ($b$) corresponds to the outcome of the measurement in the first time step (second time step) of the $j$-th  repetition and  $n_{(x,y)}$ is the total number of repetitions of the respective measurement sequence. Note that as before $n$ denotes here the number of random variables which corresponds to $n=\sum_{x,y} n_{(x,y)}$. With this it can be easily seen that due to the fact that $ \alpha_{(x,y)}^{(a,b)}\in \{0,1\}$  the sample mean corresponds to $\sum_{a,b,x,y} \alpha_{(x,y)}^{(a,b)} f_{(x,y)}^{(a,b)}$, where $f_{(x,y)}^{(a,b)}$ are the frequencies obtained in the experiment for the corresponding sequence of measurements and outcomes. Hence, it holds for the quantities $\BB_{i}^{\rm exp}$
\begin{equation}\label{hoeffdBB}
{\rm Prob}[ |\BB_{i}^{\rm exp}-\mathbb{E}(\BB_i)|\geq t]\leq 2\, {\rm exp}\left[-\frac{2t^2}{\sum_{x,y}1/n_{(x,y)}}\right].
\end{equation}
The confidence intervals are then obtained as follows. We demand that the probability for $\BB_{i}^{\rm exp}$ to deviate from $\mathbb{E}(\BB_i)$ by at most $t$ is larger or equal than $\gamma$, i.e.,
\begin{align}\nonumber
&{\rm Prob}[ \BB_{i}^{\rm exp}\in[\mathbb{E}(\BB_i)- t,\mathbb{E}(\BB_i)+t]]=\\&1-{\rm Prob}[ |\BB_{i}^{\rm exp}-\mathbb{E}(\BB_i)|\geq t]\geq \gamma,\label{eqgamma}
\end{align}
and determine $t$ as a function of $\gamma$.
The values of $t$ and $\gamma$ for which  the bounds in Eq. (\ref{hoeffdBB})  and  Eq. (\ref{eqgamma}) on ${\rm Prob}[ |\BB_{i}^{\rm exp}-\mathbb{E}(\BB_i)|\geq t] $ coincide fulfill
\begin{equation}
2 \,{\rm exp}\left[-\frac{2t^2}{\sum_{x,y}1/n_{(x,y)}}\right]=1-\gamma.
\end{equation}
This is equivalent to 
\begin{equation}
t=\sqrt{-\ln \left(\frac{1-\gamma}{2}\right)\sum_{x,y}\frac{1}{2\,n_{(x,y)}}}.
\end{equation}
Therefore,  the value of $t$ that defines a confidence interval of $68$ \%, can be calculated by using the equation above with $\gamma=0.68$. The corresponding confidence interval for $\mathcal{T}^{\rm exp}$ can be derived analogously.
It should be noted that such confidence intervals do not rely on any assumptions about the distribution of the random variables. Note further that these confidence intervals are not typically used in the analysis of ion trap experiments (for an exception see \cite{Tobias}).

The AoT constraints have been examined on the experimental data with validation (see Appendix D). 
In order to estimate the validity of the AoT constraints for the observed experimental data it suffices for sequences of length two to consider the conditions $\sum_b p(0b\vert x0) - \sum_b p(0b\vert x1)=0 $ for $x\in \{0,1\}$ as all other conditions are linearly dependent on this one. We performed a likelihood-ratio test (see e.g. \cite{Tobias} for more details) to evaluate how likely the observed violation (or higher violations) is under the assumption that the AoT constraints hold true.

In order to test the AoT constraints with high precision  we repeated the experiment performing the protocol that leads to the maximum value of the quantity $\BB_2$ with more than 12 000 repetitions per measurement sequence. Using the likelihood-ratio test we obtain for this data a statistical significance of the violation equivalent to $3.5$ standard deviations which shows that for longer runs of the experiment apparent signaling with respect to the past can be observed. 

This effect can be explained as follows. The experiment is performed in a cycle of all measurement combinations (0 0, 0 1, 1 0, 1 1). Such a cycle is then repeated more than 12000 times. This implies that each measurement always follows all measurements of the previous cycle and the measurement is affected by the measurements of the previous cycle and each cycle before. For example the pulse fidelity (due to loading of capacitors in the electrical set-up) can change depending on the sequence step. The extent of this influence is so small, that it has no significance for smaller data sets but it accumulates for long running times of the experiment. In order to examine the validity of the AoT constraints for the sequences of length 3 we consider the 14 linearly independent constraints and perform a likelihood-ratio test.

\end{document}